\documentclass[sigconf]{acmart}
\AtBeginDocument{%
  }

\setcopyright{acmlicensed}
\copyrightyear{2025}
\acmYear{2025}
\setcopyright{rightsretained}
\acmConference[CSCW Companion '25]{Companion of the Computer-Supported Cooperative Work and Social Computing}{October 18--22, 2025}{Bergen, Norway}
\acmBooktitle{Companion of the Computer-Supported Cooperative Work and Social Computing (CSCW Companion '25), October 18--22, 2025, Bergen, Norway}\acmDOI{10.1145/3715070.3749231}
\acmISBN{979-8-4007-1480-1/2025/10}

\usepackage{multirow} 
\usepackage{balance} 

\usepackage{soul}  
\usepackage{color} 
\usepackage{xspace}
\usepackage{listings}
\usepackage{enumitem}
\usepackage{hyperref}
\usepackage{appendix}
\newcommand{\eg}{{\it e.g.,\ }}

\usepackage{booktabs}
\definecolor{oxfordblue}{rgb}{0.0, 0.13, 0.28}
\definecolor{harvardcrimson}{rgb}{0.79, 0.0, 0.09}
\definecolor{dartmouthgreen}{rgb}{0.05, 0.5, 0.06}
\definecolor{princetonorange}{rgb}{1.0, 0.56, 0.0}
\definecolor{yaleblue}{rgb}{0.06, 0.3, 0.57}
\definecolor{usccardinal}{rgb}{0.6, 0.0, 0.0}
\definecolor{uclablue}{rgb}{0.33, 0.41, 0.58}
\definecolor{msugreen}{rgb}{0.09, 0.27, 0.23}
\definecolor{cornellred}{rgb}{0.7, 0.11, 0.11}
\definecolor{pomegranate}{RGB}{192, 57, 43}
\definecolor{anti-pomegranate}{RGB}{43,178,192}
\definecolor{alizarin}{RGB}{231, 76, 60}
\definecolor{anti-belize}{RGB}{185, 41, 56}
\definecolor{belize}{RGB}{41, 128, 185}
\definecolor{peter}{RGB}{52, 152, 219}
\definecolor{green}{RGB}{22, 160, 133}
\definecolor{anti-green}{RGB}{160,22,118}
\definecolor{turquoise}{RGB}{26, 188, 156}
\definecolor{pumpkin}{RGB}{211, 84, 0}
\definecolor{anti-pumpkin}{RGB}{0,22,211}
\definecolor{carrot}{RGB}{230, 126, 34}
\definecolor{wisteria}{RGB}{142, 68, 173}
\definecolor{anti-wisteria}{RGB}{99,173,68}
\definecolor{amethyst}{RGB}{155, 89, 182}
\definecolor{nephritis}{RGB}{39, 174, 96}
\definecolor{anti-nephritis}{RGB}{174,39,117}

\newcommand{\pzh}[1]{{\color{black} #1}}

\newcommand{\lxe}[1]{{\color{black} #1}}

\begin{document}

\title{``Here Comes the Makeup Tutorial You Asked For!'': Exploring Communication Strategies and Viewer Engagement in Beauty Videos on Rednote}



\author{Xueer Lin}
\affiliation{%
  \institution{Sun Yat-sen University}
  \city{Zhuhai}
  \state{Guangdong}
  \country{China}}
\email{linxer6@mail2.sysu.edu.cn}

\author{Chenyu Li}
\affiliation{%
  \institution{Sun Yat-sen University}
  \city{Zhuhai}
  \state{Guangdong}
  \country{China}}
\email{lichy235@mail2.sysu.edu.cn}

\author{Yuhan Lyu}
\affiliation{%
  \institution{Sun Yat-sen University}
  \city{Zhuhai}
  \state{Guangdong}
  \country{China}}
\email{lvyh26@mail2.sysu.edu.cn}

\author{Zhicong Lu}
\affiliation{%
  \institution{George Mason University}
  \city{Fairfax}
  \state{Virginia}
  \country{USA}}
\email{zlu6@gmu.edu}

\author{Zhenhui Peng}
\authornote{Corresponding author.}
\affiliation{%
  \institution{Sun Yat-sen University}
  \city{Zhuhai}
  \state{Guangdong}
  \country{China}}
\email{pengzhh29@mail.sysu.edu.cn}

\begin{abstract}
\pzh{
More and more people, especially females, create and view beauty videos covering topics like makeup tutorials and vlogs in social media platforms. 
Understanding the communication strategies that creators used in these videos and how their effects on the viewers' engagement can help spread beauty knowledge.
By coding 352 beauty videos in Rednote, this study presents a comprehensive taxonomy of communication strategies used by the creators, such as using home as the video background and displaying makeup effect when starting narrative at the beginning. 
We further label and computationally classify six categories of comments that reveal viewers' engagement with beauty videos. 
The regression analyses reveal the effects of the beauty video communication strategies on viewers' engagement, e.g., calling viewers to take action in the end tend to attract more comments that debate the product's efficacy. 
We discuss insights into fostering creation of beauty videos and communication of beauty knowledge. 
}

\end{abstract}

\ccsdesc[500]{Human-centered computing~Empirical studies in HCI}



\keywords{Beauty video, Rednote, Communication strategy, Viewer engagement}


\maketitle
\section{Introduction}
\pzh{
Beauty videos, covering topics like makeup tutorials and beauty product sharing, become increasingly popular in social media platforms like Rednote (Xiaohongshu in Chinese), Tiktok, and Youtube \cite{mei2024influence}. 
To attract viewers' attention and gain reputation (\eg followers) in the platform, creators need to adopt proper strategies to communicate the content and goal of the beauty videos, such as displaying makeup effects or summarizing the video content at the beginning. 
Proper usage of communication strategies can enhance the viewing experience and attract more comments, likes, and sharing from viewers, forming an active beauty community \cite{yap2024influence}. 
Apart from the benefits for creators, understanding the communication strategies of beauty videos can help viewers locate effective and interesting content, fostering informal learning of beauty knowledge. 
}

\pzh{
Prior researchers in CSCW and HCI have quantitatively understood various types of created content and members' interaction in social media. 
For example, they have explored the impact of Bilibili's scientific video dissemination strategies (\eg use visual representations of scientific concepts) \cite{zhang2023understanding} on viewers' engagement, identified the strategies (\eg use linguistic cues to indicate the impact of the work) in asking research-sensemaking questions   in community Q\&A websites \cite{he2024engage}, and investigated the hashtag usage strategies (\eg use \#BabySupplementalFood) for audience control in RedNote \cite{wan2025hashtag}. 
In the beauty domain, existing research has identified communication characteristics at both the technical (\eg special effects, video subtitles) and the informational levels (\eg narrative strategies \cite{choi2017giving}, speech acts \cite{chen2023youtubers}, influencers' discourse structures \cite{lee2022study}). 
However, there lacks a comprehensive taxonomy of the beauty videos' communication strategies and quantitative studies on how these strategies may affect the viewers' commenting behaviors. 
Specifically, two research questions (RQs) are under-addressed: 
\textbf{RQ1) what communication strategies are used in beauty videos}; and \textbf{RQ2) how would these communication strategies affect the numbers of comments (behavioral engagement), as well as the expression of attitude (emotional engagement) and discussion of beauty knowledge (cognitive engagement) in these comments under the videos?} 
}

\pzh{
To this end, we collect 352 beauty-related videos from Rednote, a platform with over 300 million monthly active users, where beauty content consistently engages a growing amount of audience \footnote{\url{https://www.qian-gua.com/blog/detail/2898.html}}. 
Two researchers employ qualitative content analysis to identify the communication strategies used in these videos and develop a taxonomy combining visual elements, narrative strategies in the videos' beginning / main / ending stages, and communicator characteristics.
For example, we find that beauty videos commonly use speaking to the camera as a video style, use expressive speech acts (\eg \textit {``This makeup is so beautiful!''}) in the main part of the video, and display the makeup effect before ending the video. 
To address RQ2, we randomly select 2,000 user comments under the collected videos and label these comments  manually coding them into categories of ``Question'', ``Attitude'', ``Storytelling'', ``Socialization'', ``Discussion'', and ``Critique''. 
A BERT-based deep learning classifier is then trained on this labeled dataset and extended to classify all comments of the collected 352 videos. 
We apply regression analyses to examine how different strategies of the videos impact viewers' behavioral, emotional, and cognitive engagement. 
The results suggest that greeting at the beginning and calling audience to take actions in the ending of beauty videos can attract more comments from viewers, using humorous tone and inserting text and image can help the videos receive more comments that express viewers' attitude, and using directive speech at the beginning and calling actions in the ending would encourage more comments that question, discuss, tell a story about, or criticize the videos. 

In summary, this paper contributes a taxonomy of communication strategies used in beauty videos, quantitative findings on how these strategies affect viewers' engagement, and insights into fostering beauty video creation. 
}

\section{Method}

\subsection{Data Collection}
\label{sec:data_collection}

To collect data, we adopt methods from previous studies \cite{wan2025hashtag}, using MediaCrawler\footnote{\url{https://github.com/NanmiCoder/MediaCrawler}} to gather 200 beauty-related videos and their metadata (e.g., likes, comments) under the ``Beauty Video'' tag. We then extract and analyze the associated hashtags, identifying the top three: ``Makeup Sharing'', ``Makeup without Cause'', and ``Makeup Today''. Another 150 videos under these hashtags are collected, resulting in a final set of 429 unique videos after removing duplicates. In addition, we crawl first-level user comments.
%
To ensure dataset reliability, we exclude videos with fewer than 50 comments, resulting in a final dataset of 352 videos and a total of 120,770 associated comments.
The video durations range from 5.87 seconds to 1185.75 seconds, with an average length of 188.21 seconds.
We categorize the videos into six types: \textit{Makeup Tutorial} (49.72\%), \textit{Makeup Vlog} (27.84\%), \textit{Makeup Change} (6.53\%), \textit{Other} (6.53\%), \textit{Single Makeup Effect Display} (5.11\%), and \textit{Product Sharing} (4.26\%).
Video likes ranged from 179 to 378,000 ($M = 18,961.07$), and comments ranged from 50 to 8,409 ($M = 343.10$).
\subsection{Data Analysis}
\label{sec:data_analysis}

\begin{figure*}[htbp]
    \centering
    \includegraphics[width=\textwidth]{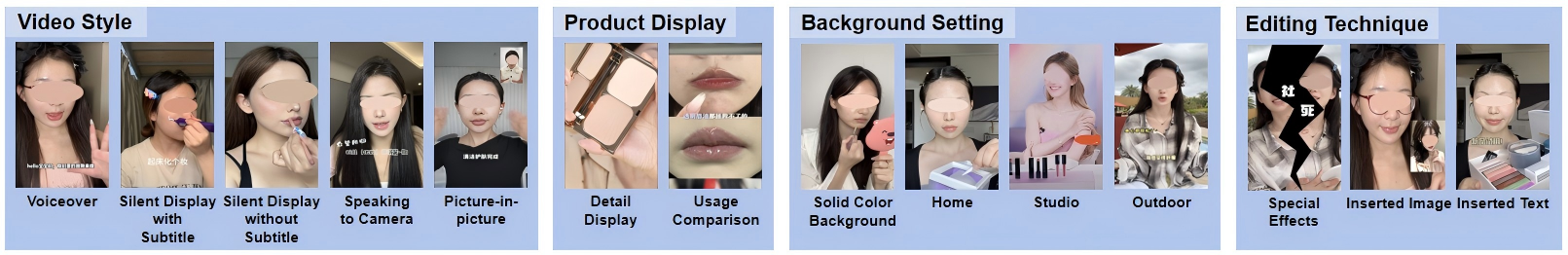}
    \caption{
    \pzh{Examples of the visual strategies used in the beauty videos}
    }
    \label{fig:combined}
\end{figure*}

To address RQ1,
two coders collaboratively develop the taxonomy through iterative discussion (see \autoref{tab:beauty_video_taxonomy}), coding 50 randomly selected videos. 
The video visual strategy example diagram is shown in the \autoref{fig:combined}. 
\lxe{Specially, we manually divide each video into three sections: beginning, main, and end.
The beginning section is determined by evaluating whether the first one minute include elements intended to capture viewer attention, such as greetings, previews, or visual highlights. The end section is identified by examining whether the last one minute contain content such as summaries, viewer interaction prompts, or calls to action. For videos lacking clear indicators, segment boundaries are adjusted based on narrative cues and visual transitions.}
Coding consistency is evaluated with Cohen's Kappa, which yields an average score of 0.80. The highest consistency in influencer professionalism and gender ($K=1.00$), while the lowest is the Beginning Speech Act ($K=0.66$).

\begin{table*}[htbp]
\small
\centering
\caption{Taxonomy of Beauty Video Communication Strategies: This table categorizes beauty video strategies into visuals, narratives, and communicators, with percentages indicating the prevalence of different values within each feature. For strategies labeled with ``Beginning/Main/Ending,'' the values in parentheses (e.g., (7.67\% / 7.39\% / 5.40\%)) represent the proportion of the strategy's use in the video’s introduction, main content, and conclusion, respectively.}
\begin{tabular}{p{1.9cm}p{4.8cm}p{4.0cm}p{5.7cm}}
\toprule
\textbf{Category} & \textbf{Feature} & \textbf{Definition} & \textbf{Value} \\
\midrule
\multirow{4}{*}{Visual} 
& Video Style \cite{morcillo2015typologies,welbourne2016science} & Presentation mode of the video & \textit{Speaking to Camera} (38.08\%), \textit{Silent Display without/with Subtitle} (27.27\%, 24.15\%), \textit{Voiceover} (12.20\%), \textit{Picture-in-picture} (1.14\%) \\
& Product Display \cite{xu2021influence} & How beauty products are shown & \textit{Detail Display} (75.85\%), \textit{None} (23.58\%), \textit{Usage Comparison} (1.42\%) \\
& Background Setting \cite{zhang2018evoking} & Filming Environment & \textit{Home} (80.40\%), \textit{Solid Color Background} (10.80\%), \textit{Studio} (4.55\%), \textit{Outdoor} (4.55\%) \\
& Editing Technique \cite{choi2017giving, kim2022study} & Editing elements used & \textit{Quick Cuts} (74.43\%), \textit{BGM} (74.15\%), \textit{Inserted Text} (22.44\%), \textit{Special Effects} (21.59\%), \textit{Inserted Image} (16.76\%), \textit{None} (2.27\%) \\
\midrule
\multirow{5}{*}{Narrative} 
& Beginning Narrative Technique \cite{wilkens2003role} & Narrative techniques used at the beginning to introduce the video or engage viewers &  \textit{Greeting} (29.26\%),  \textit{Personal Story} (8.81\%),  \textit{Scene Setup} (1.70\%),  \textit{Summary Preview} (38.07\%),  \textit{Makeup Effect Display} (30.97\%),  \textit{Suspense Question} (5.40\%),  \textit{Viewer Comment} (6.53\%),  \textit{Start Directly} (29.83\%) \\
& Beginning/Main/Ending Emotion \cite{wang2022identity, wei2023re} & Usage of exaggerated movements to evoke emotions in the audience & \begin{tabular}[t]{@{}l@{}}\textit{Exaggerated Expression} (7.67\% / 7.39\% /5.40\%)\\ \textit{Exaggerated Storytelling} (7.67\% /7.95\% /1.70\%)\\ \textit{None} (84.94\% /84.94\% /92.90\%) \end{tabular}\\
& Beginning/Main/Ending Speech Act \cite{chen2023youtubers, munaro2024does} & Types of speech behaviors used at different structural stages of a video (beginning, main, and ending) 
&\begin{tabular}[t]{@{}l@{}}\textit{Assertive} (12.78\% /47.73\% /5.11\%)\\ \textit{Directive} (4.83\% /25.28\% /3.69\%)\\ \textit{Commissive} (5.11\% /2.56\% /3.98\%)\\ \textit{Expressive} (38.92\% /36.08\% /20.74\%) \\ \textit{Quoting} (3.12\% /1.70\% /0.0\%) \\ \textit{None} (47.73\% /30.68\% /71.02\%) \end{tabular}\\
& Beginning/Main/Ending Tone \cite{he2024face} & Influencer’s tone & \begin{tabular}[t]{@{}l@{}} \textit{Neutral} (19.60\% /25.85\% /15.06\%) \\ \textit{Enthusiastic} (23.58\% /17.05\% /10.80\%)\\ \textit{Professional} (1.14\% /11.08\% /0.57\%) \\ \textit{Humorous} (5.11\% /7.10\% /0.85\%) \\ \textit{None} (51.14\% /41.19\% /72.73\%) \end{tabular} \\
& Ending Narrative Technique* & Narrative cues used to summarize, interact, or close the video & \textit{Makeup Effect Display} (74.43\%), \textit{Farewell} (20.74\%), \textit{No Ending} (12.50\%), \textit{Summary} (4.83\%), \textit{Call to Action} (4.55\%), \textit{Set Expectation} (3.69\%), \textit{Personal Story} (3.12\%), \textit{Invite Comments} (2.27\%), \textit{Blessing} (1.14\%) \\
\midrule
\multirow{2}{*}{Communicator} 
& Professionalism \cite{chong2018youtube, riboni2017between} & Whether the influencer is a professional makeup artist & \textit{Non-professional} (97.44\%), \textit{Professional} (6.53\%) \\
& Gender \cite{li2025constructing, karjo2020language,komulainen2017men} & Influencer’s gender & \textit{Female} (97.44\%), \textit{Male} (2.56\%) \\
\bottomrule
\end{tabular}
\footnotesize{* These features were newly found from the beauty video on Rednote.}
\label{tab:beauty_video_taxonomy}
\end{table*}

To address RQ2, this study analyzes beauty video communities using the classification framework from \cite{zhang2023understanding}, with modifications based on the actual comments under beauty videos. We categorize user comments into six categories as \autoref{tab:combined_comment_table}: \textit{Discussion}, \textit{Narration}, \textit{Question}, \textit{Criticism}, \textit{Attitude}, and \textit{Socialization}. For annotation reliability, two authors independently label 1,000 comments ($Cohen's\ Kappa = 0.89$), followed by additional annotations of 1,000 comments. 
To improve classification accuracy, we further subdivide each main comment category into sub-labels for annotation.
A fine-tuned Chinese BERT model (80\% training, 20\% validation) \cite{devlin2019bert} processing 2,000 samples is used to generate single-label predictions. We label 120,770 valid comments labels using this classifier with 98.14\% accuracy.
Three regression models are developed to examine how communication strategies relate to three engagement dimensions: (1) \textbf{behavioral engagement}, \lxe{measured by the total number of first-level comments, encompassing all six comment categories; }
(2) \textbf{emotional engagement}, indicated by the proportion of \textit{Attitude} comments (e.g., \textit{``This foundation matches my skin tone perfectly!''}); and (3) \textbf{cognitive engagement}, reflected in the proportion of knowledge-intensive interactions (\textit{Discussion, Storytelling, Question, Critique}), particularly those involving product efficacy and technical analysis.

\begin{table*}[htbp]
\centering
\small
\caption{Beauty Video Comment Category. We categorize the comments on beauty videos on Rednote into six categories. Each category's definition, example, proportion, and sub-label are as follows.}
\label{tab:combined_comment_table}
\begin{tabular}{p{1.2cm}p{2.9cm}p{3.4cm}p{3.4cm}c}
\toprule
\textbf{Category} & \textbf{Definition} & \textbf{Example (Translated)} & \textbf{Sub-label} & \textbf{Proportion (\%)} \\
\midrule
Question & Makeup or product details queries & \textit{``Which lip gloss shade complements fair skin best?''} & 
\begin{tabular}[t]{@{}l@{}} Non-beauty/Beauty Inquiry,   \\ 
 Other Questions\end{tabular} 
& 32.67 \\
\addlinespace[0.7ex]
Attitude & Emotional evaluations of content quality & \textit{``This makeup tutorial is exceptionally well-executed!''} & 
\begin{tabular}[t]{@{}l@{}} Content/Influencer Appreciation, \\ Content/Influencer Disapproval\end{tabular} & 31.33 \\
\addlinespace[0.7ex]
Storytelling & Personal cosmetic experiences & \textit{``The white particles in my hair aren't dandruff...''} & 
\begin{tabular}[t]{@{}l@{}}Personal Experience Sharing, \\ Other Narratives\end{tabular} & 11.24 \\
\addlinespace[0.7ex]
Socialization & Platform-specific interaction rituals & \textit{``Content creator please notice me!''} & 
\begin{tabular}[t]{@{}l@{}}Polite Greeting,\\ Share with Friends\end{tabular} & 15.28 \\
\addlinespace[0.7ex]
Discussion & Technical suggestions for improvement & \textit{``Try androgynous contouring instead..''} & 
\begin{tabular}[t]{@{}l@{}}Beauty-related Discussion,\\
Influencer-related Discussion\end{tabular} & 9.28 \\
\addlinespace[0.7ex]
Critique & Critical analysis of authenticity & \textit{``Heavy filters distort the actual makeup effects''} & 
\begin{tabular}[t]{@{}l@{}}Video Production/Product Criticism, \\ Other Criticisms\end{tabular} & 0.20 \\
\bottomrule
\end{tabular}

\end{table*}
\section{Findings}
\subsection{Communication Strategy of Beauty Videos (RQ1)}
The proportion of videos using the video strategy is shown in the ``Value'' column of the \autoref{tab:beauty_video_taxonomy}.
\textbf{In terms of visual strategies, beauty videos commonly use the speaking to the camera as a video style, display product detail, and take dynamic video editing techniques such as quick cuts and well-matched background music.}
The speaking to camera style is widely used in beauty videos, allowing influencers to interact directly with the audience and establish a personal connection \cite{chong2018youtube}. 
Additionally, some beauty videos present content without speech, occasionally supplemented by subtitles. This approach aligns with the ``Immersive Makeup'' trend, which prioritizes visual aesthetics and the effective presentation of cosmetic products.
Moreover, detailed product showcases, such as influencers demonstrating foundation texture during application, are often included. Quick cuts and well-matched background music are also incorporated in video editing to keep the content dynamic. These editing techniques fit the Rednote platform's temporary attention characteristics and maintain the persistence of viewing behavior \cite{ten2015like}. Text (e.g., product link for purchase) and images (e.g., meme) are frequently inserted to add details and increase the video’s entertainment value. 

\textbf{In terms of narrative technique and emotion evoking, beauty videos commonly summarize content at the beginning, display makeup effects at the end, and avoid emotional triggers.}
The beginning often summarizes the video content (e.g., \textit{``Today, I'll share a quick makeup look for the morning class outing''}), helping the audience clarify the theme. At the same time, some videos also use strategies to show the effect of makeup at the beginning, which can quickly attract the audience's attention, and this is also a common strategy at the end of the video. Further research reveals that the instant visual impact formed by the makeup effect display at the beginning or end of the video can effectively stimulate emotional reactions such as amazement and pleasure. This arrangement is precisely consistent with the fast-paced consumption characteristics of short videos \cite{wilkens2003role}. Most beauty videos do not use emotional triggers (e.g., exaggerated storytelling), indicating a more rational presentation.
 
\textbf{In terms of speech and tone, beauty videos commonly use expressive speech and an enthusiastic tone at the beginning, followed by assertive statements and a neutral tone in the main content.}
Nearly half of the videos begin without words, emphasizing the visuals. When speech is used, expressive phrases like \textit{``This makeup is so beautiful''} foster emotional connection. For the main content, assertive speech (e.g., \textit{ ``This lipstick is perfect for me''})  and expressive speech acts (e.g., \textit{``I’m so glad that I got my salary today''}) are common. However, most videos end without speech. In addition, an enthusiastic tone is used at the beginning to build rapport, while a neutral tone in the body helps viewers focus on the explanation.

\textbf{In terms of communicators, most beauty videos are created by non-professional makeup enthusiasts, primarily featuring female communicators.}
The vast majority of beauty video creator is non-professional makeup enthusiasts, emphasizing personal experience rather than expert authority. 
While professional voices are still a minority, their joining can be strategically positioned to increase the credibility of beauty videos.
This is consistent with the overwhelming female communicator base, indicating a strong gender identity in content creation.
While gender identity is implicit in the predominance of female creators, it also influences how beauty norms are expressed and interpreted on the platform. Prior research on beauty-related user-generated content has noted the frequent use of gendered language \cite{abidin2016aren}, such as expressions of intimacy and solidarity among female users (e.g., “sister” or “girl power”), which often reinforce themes of self-care, confidence, and mutual support. Although our study does not include a systematic analysis of comment content, these patterns suggest that gendered narratives may play a key role in shaping both content creation and audience engagement, offering a valuable direction for future research.

\subsection{The Impact of Communication Strategies on Viewers' Engagements (RQ2)}

\begin{table*}[htbp]
\small
\caption{Regression analysis of video communication strategies and user engagement: This table analyzes the facilitating/inhibiting effects of specific strategies on the behavioral, emotional, or cognitive engagement.}
\label{tab:regression}
\begin{tabular}{llllll}
\toprule
Engagement & Feature (Value) & Coef. & Std. & t & p\\
\midrule
\multirow{4}{*}{Behavioral} & Video Style (Without Subtitle) & -0.015 & 0.002 & -9.85 & < 0.001*** \\
 & Beginning Narrative Technique (Greeting) & 0.008 & 0.001 & 7.79 & < 0.001*** \\
 & Ending Narrative Technique (Call to Action) & 0.012 & 0.002 & 7.54 & < 0.001*** \\
 & Main Speech Act (Assertive) & -0.007 & 0.001 & -4.86 & < 0.001*** \\
\midrule
\multirow{3}{*}{Emotional} & Main Tone (Humorous) & 0.057 & 0.008 & 7.62 & < 0.001*** \\
 & Editing Technique (Inserted Text) & 0.030 & 0.004 & 7.29 & < 0.001*** \\
 & Editing Technique (Inserted Image) & 0.029 & 0.005 & 6.30 & < 0.001*** \\
\midrule
\multirow{3}{*}{Cognitive} & Main Speech Act (Expressive) & -0.038 & 0.005 & -8.15 & < 0.001*** \\
 & Beginning Speech Act (Directive) & 0.085 & 0.011 & 7.54 & < 0.001*** \\
 & Ending Narrative Technique (Call to Action) & 0.085 & 0.008 & 10.47 & < 0.001*** \\
\bottomrule
\end{tabular}
\end{table*}
\autoref{tab:regression} presents the results of multiple linear regression with communication strategies as independent variables and three viewer engagement metrics as dependent variables.
\textbf{Greeting at the beginning and calling
audience to take actions in the ending of beauty videos can attract more comments from viewers (Behavioral Engagement).}
The \textit{Silent Display Without Subtitles} strategy (e.g., recording a makeup process with no spoken words and subtitles) significantly reduces user interaction, aligning with the ``Immersive Makeup'' trend, which emphasizes visual appeal over traditional interaction to increase likes.
Greeting the audience at the beginning enhances social presence and fosters engagement through actions like liking, commenting, and sharing \cite{kim2024value, chen2023hardcore}. A call to action at the end also boosts engagement. However, assertive statements (e.g., “It is suitable for oily skin.”) may hinder behavioral engagement and discussion due to their absolute nature. While such claims reduce cognitive ambiguity and strengthen product persuasion, they can also limit openness to interaction, thereby reducing audience feedback \cite{hock2023selling, huang2023and}.

\textbf{The combination of entertaining tones and informative visual cues can enhance the audience's emotional expression (Emotional Engagement).}
Specifically, the humorous tone of the make up process stimulates emotional resonance \cite{bernad2023multimodal}, and the insertion of text (e.g., product name) and images (e.g., netizens' comments) enhances information intuition, which together enhance both the entertainment value and cognitive clarity for the audience. 
The systematic combination of multiple strategies ultimately validates the empowering effect of visual narrative and emotional arousal mechanisms on the vitality and attractiveness of content on the Rednote platform \cite{ten2015like}.


\textbf{Speech acts and call-to-action cues in beauty videos serve as critical touchpoints that foster deeper viewer reflection on beauty knowledge (Cognitive Engagement).}
Directive speech act at the beginning of the video (e.g.,\textit{``Come try this makeup.''} ) shows significant positive feedback, building viewing commitment through immediate interaction and setting the stage for subsequent messaging. 
It is worth noting that \textit{expressive speech act} (e.g.,\textit{``I'm super excited.''} ) can strengthen the emotional tone through emotional expression, but its negative correlation with cognitive engagement suggests that the creators need to balance the emotional intensity with the information value \cite{hock2023selling, huang2023and}. 
The ending is achieved through a call to action for the viewer (e.g., \textit{ ``Everyone can take action according to the tutorial!''}) that stimulates the viewers' thinking behavior, which indirectly enhances their cognitive engagement and thus strengthens the internalization and understanding of the message \cite{kim2024value, chen2023hardcore, tyree2016making}. 

\section{Discussion and Conclusion}

\pzh{
This work develops a taxonomy for common communication strategies used in 352 Rednote beauty videos and quantitatively reveals these strategies' impacts on the viewers' commenting behaviors. 
The findings complement CSCW research on understanding user-generated content and members' interactions in social media, providing insights into fostering the creation of beauty videos and learning of beauty knowledge. 
For example, using our annotated videos as the starting point, researchers can design posting support tools \cite{peng2020exploring,zhang2025mentalimager} that recommend example video clips to creators when they want to adopt certain strategies. 
Based on the results of our regression analyses, the tools can recommend strategies like using directive speech act at the beginning and calling viewers to take actions in the ending if the creators want to receive more comments that discuss the beauty knowledge. 
A concrete usage scenario might involve an integrated tool within a video editing platform, where a creator receives real-time suggestions during the scriptwriting or editing process. When the tool detects that the creator is preparing a tutorial-style beauty video, it could recommend inserting phrases or clip segments associated with higher knowledge-sharing engagement. 
Additionally, to provide an engaging learning experience to viewers who want to gain beauty knowledge, researchers can develop tools that recommend beauty videos which adopt communication strategies positively link to viewers' behavioral engagement.  
}


This study also has several key limitations. First, it excludes factors like the influencer's follower count, which could provide additional insights into viewer engagement \cite{Ttafesse2021followers, shi2024study}. 
Future research should consider these factors for a more comprehensive analysis. Second, the dataset contains a limited number of videos in certain categories, such as product-sharing videos (4.26\%), which may impact the robustness of the findings. Expanding the dataset with more diverse video types would improve the model's generalization performance. Third, we excluded videos with fewer than 50 comments to ensure sufficient engagement data for analysis. However, this may exclude niche creators or newly posted content, potentially overlooking emerging trends. Future work could consider dynamic thresholds or incorporate temporal engagement measures to address this.
Lastly, while this study focuses on Rednote, a platform with limited prior research, it does not compare content or audience behavior across other major video platforms such as YouTube or TikTok. Future work could examine cross-platform and cultural differences to better understand how contextual factors shape beauty content and viewer engagement.

\section*{Acknowledgments}
This work is supported by the Young Scientists Fund of the National Natural Science Foundation of China (NSFC) with Grant No. 62202509, NSFC Grant No. U22B2060, and the General Projects Fund of the Natural Science Foundation of Guangdong Province in China with Grant No. 2024A1515012226.
\balance
\bibliographystyle{ACM-Reference-Format}
\bibliography{main}

\appendix

\end{document}